\documentclass[fleqn,10pt]{wlscirep}
\usepackage[utf8]{inputenc}
\usepackage[T1]{fontenc}
\usepackage{siunitx}
\title{Demultiplexing through a multimode fiber using chip-scale diffractive neural networks}

\author[1,+*]{Qian Zhang}
\author[2,+]{Haoyi Yu}
\author[1]{Jie Zhang}
\author[1]{Yuedi Zhang}
\author[2]{Chao Meng}
\author[1]{Jiali Sun}
\author[1]{Yu Miao}
\author[2]{Qiming Zhang}
\author[2,*]{Min Gu}
\author[1,*]{Juergen W Czarske}

\affil[1]{Laboratory of Measurement and Sensor System Technique, TU Dresden, Helmholtzstrasse 18, 01069, Dresden, Sachsen, Germany}
\affil[2]{School of Artificial Intelligence Science and Technology, University of Shanghai for Science and Technology, Shanghai 200093, China}

\affil[*]{juergen.czarske@tu-dresden.de, qian.zhang@tu-dresden.de, gumin@usst.edu.cn}

\affil[+]{these authors contributed equally to this work}

\begin{abstract}

In today's information age, advanced fiber optic transmission technology is of paramount importance. Multimode fibers (MMFs) using space-division multiplexing (SDM) are promising for improved transmission capacity, connection flexibility, and security of data. However, the complex transmission characteristics of MMFs significantly hinder precise mode demultiplexing. Conventional approaches, including holographic measurements, phase retrieval algorithms, photonic lanterns, and multiplane light conversion, are limited by system complexity, size, and flexibility. In this paper, we demonstrate for the first time a purely optical, chip-scale AI solution for high-mode isolation, speed-of-light demultiplexing of MMF modes using a three-dimensional diffractive neural network (DNN). The DNN is trained with synthetic modal data and fabricated using two-photon nanolithography. It features a compact size of 
\SI{120}{\micro\metre} $\times $\SI{120}{\micro\metre} $\times $\SI{80}{\micro\metre} and a diffractive structure size of \SI{1}{\micro\metre\squared} for the neurons at the hidden layers of the network. Experimentally, the DNN demultiplexer achieves a relative demultiplexing accuracy of over 80\%. The AI approach of DNN allows for flexible design and overcomes the size and performance limitations of digital-optical demultiplexers. This work paves the way for compact, low-latency optical processors for high-performance demultiplexers and enables scalable, chip-integrated solutions for next-generation fiber optic networks.

\end{abstract}
\begin{document}

\flushbottom
\maketitle
%\thispagestyle{empty}

% \noindent Please note: Abbreviations should be introduced at the first mention in the main text – no abbreviations lists. Suggested structure of main text (not enforced) is provided below.

\section*{Introduction}

Optical fibers, including single-mode fibers (SMFs) and multicore SMFs~\cite{nakajima2017multi}, have been fundamental to the development of modern information technologies due to their inherent advantages of long-distance reliability, low attenuation, minimal interference, high security, and large information-carrying capacity~\cite{ballato2017glass}. As demand for data throughput continues to grow exponentially with cloud computing, mobile and edge computing, and the Internet of Things (IoT)~\cite{ballato2017glass,machuca2022long, ellis2017performance,winzer2018fiber}, fiber optics have become increasingly critical for addressing the bandwidth limitations of conventional communication technologies.

\begin{figure}[t!]
    \centering
    \includegraphics[width=0.8\textwidth]{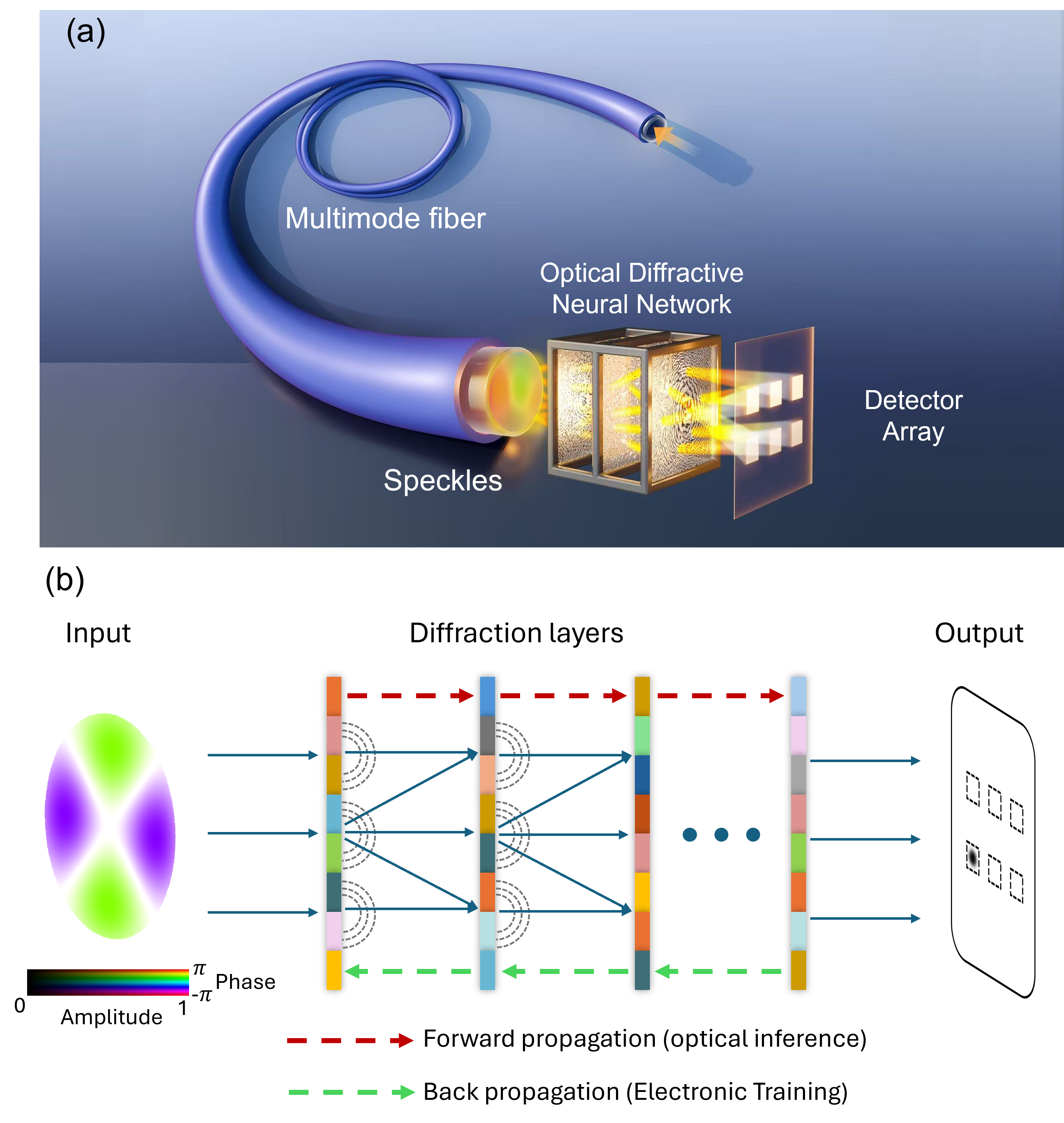}
  \caption[Flowchart of DNN]{
  (a) Schematic illustration of demultiplexing MMF using DNN. Pure eigenmodes are launched into the MMF, while the output becomes a speckle field due to coherent superposition. A six-mode example is shown here. The DNN separates the speckle field into individual mode channels based on their respective modal contributions. The modal energy can then be directly measured using photodetectors.
  (b) Schematic of a DNN-based mode demultiplexer. The input of the DNN is the superposed complex light field emerging from MMF. The DNN consists of several diffractive layers that shape the input into the corresponding detector area. 
  The detector layer contained 6 different detection regions corresponding to 6 different modes. The relative intensities of light in different detection regions can be measured using either normal light detectors or single-photon detectors, facilitating the transmission of high-dimensional information. The forward propagation of the DNN is done by ASM, and the back propagation is used to train the diffractive layers. 
 }
  \label{fig:introduction_overview_md}
\end{figure}

To meet these growing demands, optical fiber-based space-division multiplexing (SDM) technologies have been widely explored, where multiple linearly overlapping optical modes are (de-)multiplexed to enhance information capacity~\cite{richardson2013space,carpenter2012degenerate}. Research in SDM has intensified in recent years~\cite{essiambre2010capacity, shibahara2019dmd}, as SMF-based networks, especially those employing wavelength-division multiplexing (WDM), are more and more reaching their nonlinear Shannon limits~\cite{shannon2006communication,ellis2009approaching, essiambre2010capacity,ellis2017performance}, due to the intrinsic single-mode nature of SMFs.
%In contrast, 
Multimode fibers (MMFs), which support the simultaneous propagation of multiple spatial modes, are indispensable in applications such as high-power fiber lasers~\cite{zervas2014high,rothe2025wavefront}, secure communication~\cite{rothe2023securing}, and quantum key distribution (QKD) systems \cite{pirandola2016physics}. MMFs are the key component of a modern multiple-input multiple-output (MIMO) communication system.
However, their complex transmission characteristics present a significant challenge: modal crosstalk during propagation perturbs the amplitude, phase, and polarization of the transmitted light, scrambling the spatial information carried by each mode~\cite{kupianskyi2024all}, see Figure~\ref{fig:introduction_overview_md}(a). 
Coherent detection followed by advanced digital signal processing (DSP) techniques is the standard and necessary method for compensating modal crosstalk in high-capacity SDM systems~\cite{ryf2012mode, weng2019recent}.
Accurate recovery of modal weights is therefore critical for practical applications, such as digital optical phase conjugation (DOPC) for classical and quantum communication calibration~\cite{zhou2021high}, transmission matrix inversion for physical layer security over MMF~\cite{rothe2023securing}, and image transmission~\cite{butaite2022build}. However, the measurement of modal weights is inherently challenging due to the high spatial complexity of the output speckle fields. 

% digitally multiple-input multiple-output (MIMO) digital signal processing (DSP)~\cite{van2014time} 

Existing methods for mode analysis and demultiplexing of MMFs typically rely on interferometric systems~\cite{lyu2017fast} or phase-retrieval algorithms~\cite{shapira2005complete}, which substantially increase system complexity. 
All-fiber solutions, such as photonic lanterns, offer minimal-loss demultiplexing but face scalability challenges~\cite{birks2015photonic, fontaine2022photonic}.
Free-space techniques, including multi-plane light conversion (MPLC) with spatial light modulators (SLMs)~\cite{fontaine2019laguerre,liu2023broadband,kupianskyi2023high} and metasurface-based optics~\cite{liu2022multichannel, deng2022demand, zhao2025neuro},
provide higher flexibility but are either bulky or require precise nanometer-scale alignment, limiting practical integration. 
Traditional mode decomposition (MD) methods, often based on 2D intensity~\cite{an2020deep,rothe2021intensity} or interferometric measurements~\cite{lyu2017fast},
provide accurate modal weights but suffer from limited speed and high latency (<kHz)~\cite{zhang2022learning,zhang2025fpga}, while all-optical MD using SLMs is constrained by slow switching rates~\cite{gervaziev2020mode, flamm2012mode}

Optical and photonic neural networks~\cite{borhani2018learning,rademacher20221,zhang2019artificial,xu2025intelligent}, especially the development of optical diffractive neural networks (DNN), is regarded as a promising computational unit capable of processing information at the speed of light with ultra-low energy consumption~\cite{lin2018all}. Recently, various tasks have been demonstrated using DNNs, such as imaging reconstruction~\cite{yu2025all}, image recognition, linear matrix operations, logic operations and beam shaping \cite{sun2023review,sheng2022review}. 
DNNs-based on SLMs
%and bulky optics 
have been demonstrated for spatial mode manipulation, such as orbital angular momentum (OAM) multiplexing~\cite{liu2023broadband, feng2025high}, mode mapping~\cite{wang2024diffractive}, and mode converter~\cite{soma2025complete}. However, a chip-scale, ultra-compact DNN for MMF mode demultiplexing remains a challenging task, due to the fabrication capability of a multi-layer structure, and mode-mismatch between the modes in MMF and 
%bulky
optical devices.

%  However, a systematic investigation of the performance on demultiplexing superposed MMF outputs remains lacking to our knowledge.

In this work, we propose and demonstrate the all-optical, high-fidelity, and light-speed demultiplexing of modes through a typical commercial MMF using a chip-scale diffractive neural network at the telecommunication wavelength (\SI{1568}{\nano\meter}). The concept is shown in Figure~\ref{fig:introduction_overview_md}(a), where the DNN demultiplexes different eigenmodes into distinct spatial positions.
%For simplicity, we refer to this device as the DNN demultiplexer chip (DNNs).   
The DNN was modeled using the Angular Spectrum Method (ASM), and the phase distributions of the diffractive layers were optimized via back-propagation within a neural network framework.
Synthetic data of the eigenmode distributions of a MMF is used during the training process to realize the mode demultiplexing. Experimentally, a 3-layer DNN-chip with a diffractive neural size of \SI{1}{\micro\metre\squared} and a footprint of \SI{120}{\micro\metre} $\times $\SI{120}{\micro\metre} $\times $\SI{80}{\micro\metre} was fabricated using a home-built galvo-scanning two photon nanolithographic (GS-TPN) system.
With this well-trained DNN, the transmitted modes through the MMF can be demultiplexed accurately at the speed of light with a mode isolation of \SI{13.5}{\dB} between different modes. The relative error of demultiplexing can reach about $6.93\%$. 
Furthermore, compared to traditional methods, the demonstrated DNN-based demultiplexing chip can operate on broadband and full polarizations.
%mmune to the wavelengthreal-time vibrations of the MMF. 
Therefore, the 3D nanoprinted DNNs-chip is an all-optical, chip-scale demultiplexer for MMFs at the speed of light, which overturns the size and performance hurdle of conventional mode demultiplexing methods using bulky optics. 
These results demonstrate a pathway toward a compact and low-latency optical processor for future fiber-based high-capacity communication systems.
Beyond mode demultiplexing, the same strategy can also be extended to on-chip communication chips~\cite{yu2025all, yang2020terahertz}, and quantum photonics~\cite{leedumrongwatthanakun2020programmable}.

\section*{Results}

\subsection*{Design and operating principle of the optical }

To realize the DNN, we first constructed the DNN architecture in simulation, as shown in Figure~\ref{fig:introduction_overview_md}(b). 
The implementation of DNN consists of two parts: the simulation of optical light propagation between adjacent diffractive layers and the construction of a trainable model. In this work, the angular spectrum method (ASM) is used to simulate the propagation of light waves. The light field emerging from MMF is the input of the DNN, and the output is the light field modulated by the diffractive layers. 
Each diffractive layer consists of learnable pixels (neurons), which are capable of modulating the phase of the incident optical field. The modulated optical field then propagates over a certain distance and reaches the next diffractive layer for further encoding. Finally, the input light field is progressively shaped into the desired output pattern at the detector plane. This is the optical inference, which is also defined as forward propagation, see Figure~\ref{fig:introduction_overview_md}(b). In order to optimize the neuron values (learnable parameters in DNN in this work, which represent phase delay), we calculate the difference between the predictions and labels by a loss function. Afterwards, back-propagation is used to update the neuron values and minimize the error between the predictions and labels as shown in Figure~\ref{fig:introduction_overview_md}(b). 
The goal here is to train the DNN, in other words, to optimize the phase masks, to map the input to tailored detector regions.

This work presents the first application of the DNN for the demultiplexing of optical fiber modes. Distortion-free information transmission can be achieved even in the presence of mode conversions, e.g., due to fiber bending, using a DSP-based posateriori algorithm~\cite{ryf2012mode, ryf2012space}. 
% In this paper, the DNN is used for demultiplexing a MMF, which can be further employed for distortion-free information transmission, as mode conversions can be measured and corrected or exploited. 
Accordingly, the input to the DNN consists of superposed MMF output fields, while the output of DNN corresponds to the modal amplitude weights of the individual eigenmodes.
Unlike electronic neural networks, optical neural networks cannot directly produce numerical weight values. To facilitate practical readout, the DNN output is encoded in optical power, mapping the modal weights to intensity distributions that can be directly measured by photodetectors. This achieves demultiplexing, which can be used directly for data transmission with space division multiplexed communication.
In this work, two types of training datasets are generated and compared.
The first consists of the eigenmode fields, while the second contains pseudo-random modal weight combinations (PRMC) corresponding to the superposed MMF output fields.
The hyperparameters and the details of the training platform are listed in Methods. In this study, we first verified the feasibility and performance of DNN-based DEMUX in simulations, and then demonstrated it in experiments using a designed setup towards 6 modes-based information transmission.

\subsection*{Numerical Performance of DNN}

\begin{figure}[tbp]
    \centering
    \includegraphics[width=0.86\textwidth]{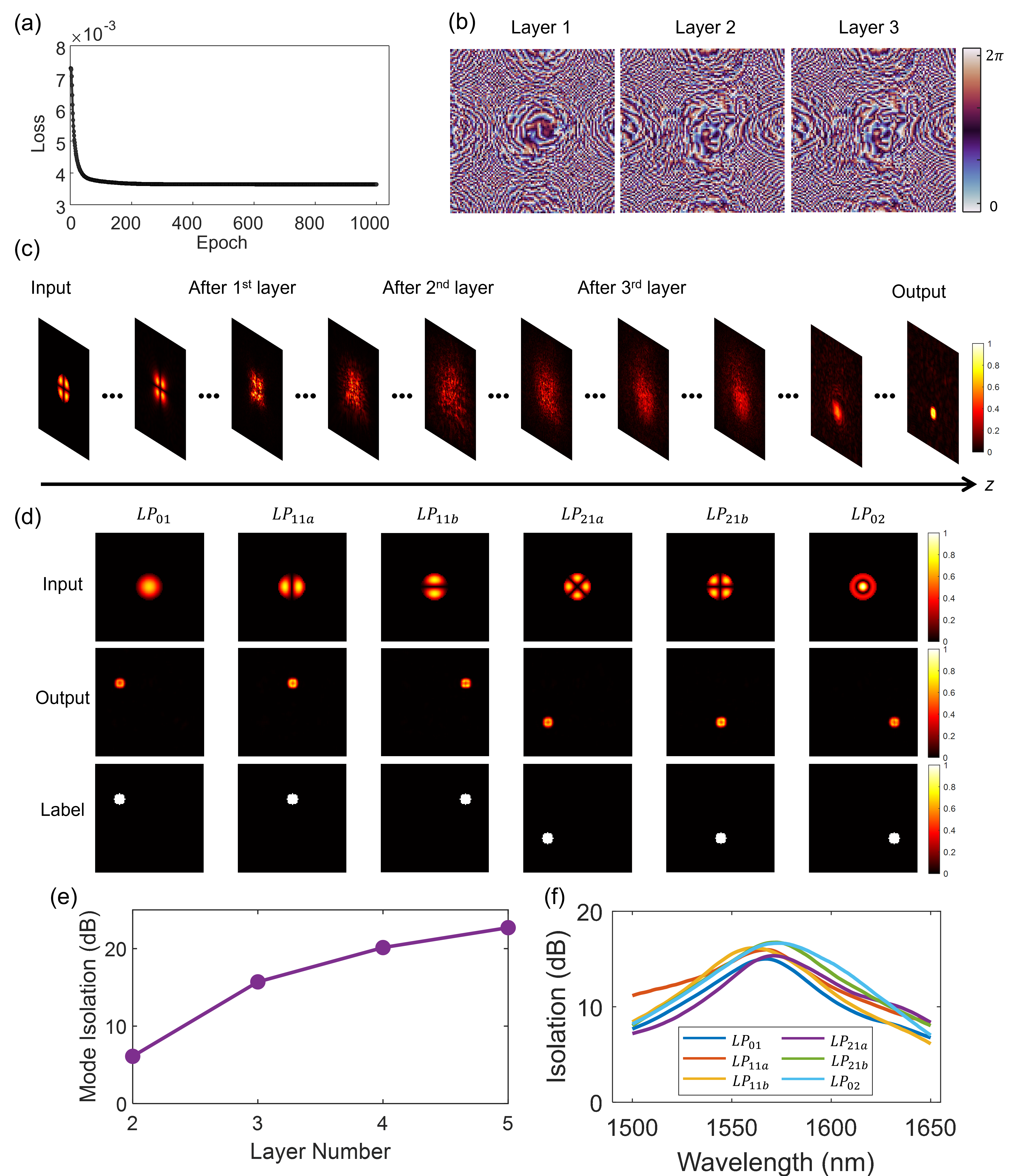}
  \caption[Visualization of the light propagation inside DNN for mode classification.]{ Performance evaluation of DNN on synthetic data.
  (a) Training process of DNN.
  (b) Phase distributions of three diffractive layers in the 3-layer DNN.
  (c) Visualization of the light propagation inside DNN for mode $LP_{11a}$. 
  (d) The predicted output field of DNN for all six modes.
  (e) Mode isolation performance under different DNN depths. Here, the DNN varies from 2 to 5 layers. 
  (f) Mode isolation across different optical wavelengths. Here, the wavelength varies from \SI{1500}{\nano\meter} to \SI{1650}{\nano\meter}.
  }
  \label{fig:results_simulation_eigenmodes_6modes}
\end{figure}

The foundational concept for this work is to electronically train the DNN and perform optical inference in the real world. We first investigate the performance of DNN in simulation. 6 LP eigenmodes, $LP_{01}$, $LP_{11a}$, $LP_{11b}$, $LP_{21a}$, $LP_{21b}$, and $LP_{02}$ are considered in this work. The complex light field distributions of these eigenmodes are calculated in the simulation based on the parameters of one commercial MMF used in the experiment.
Arbitrary optical fields can then be generated by randomly combining the modes with amplitude and phase weights for defined polarizations. Amplitude weights are L2-normalized to satisfy energy conservation, while all phase weights are randomly distributed.
The parameters of the DNN model are set to match the performance of 3D printing technology. 
The resolution of the diffractive layers in the DNN model is set to 110×110, where the physical size of each pixel is \SI{1}{\micro\metre}  corresponding to the resolution of fabrication. In total, 12 100 neurons at each hidden layer are implemented.
%Since the light field spreads out during the propagation process, absorbing padding is used here. 
Additionally, in order to avoid the wrap-around effect in ASM~\cite{matsushima2009band}, an absorption area is added during the numerical simulation, see \textbf{Figure S2}. 
The distance between the diffraction layers is set to \SI{40}{\micro\metre} to enable full connectivity via diffraction and to manufacture a compact demultiplexer unit. Details on how the distance is determined can be found in the \textbf{Supplementary Materials}. The propagation distance between the last diffractive layer and the detector is defined as \SI{120}{\micro\metre}.
The target field consists of six regions, each corresponding to a specific mode.
The intensity (i.e., pixel) value within each detection region represents the optical power of the corresponding mode.
Each target area has a size of $10\times 10$ pixels, and the detector area has the same size. The wavelength used in the simulation is \SI{1568}{\nano\meter}. 
In practical communication systems, broadband sources are typically employed. Therefore, we carried out a dedicated bandwidth test.

Two test scenarios are considered in the simulation: one is the pure eigenmodes distribution, and the other is the arbitrary mode combinations.
The performance of DNN on pure eigenmodes is tested for 2 reasons. The first is that pure eigenmode fields refer to the case where there is no crosstalk occurring during the propagation in MMF, like ultra-low crosstalk fibers~\cite{ge20196}. The second is that, considering the reversibility of the optical path, if the DNN can classify pure eigenmodes, it also means that the DNN can be used in reverse to generate different eigenmodes. More precisely, this is then a DNN-based optical multiplexer (MUX). 
In contrast, the second test scenario with arbitrary mode combinations better represents the realistic output fields of a MMF due to modal crosstalk.
During dataset generation, superposed optical fields were constructed using randomly assigned amplitude and phase weights.
As the modal energy ratio depends on the amplitude weights, the relative amplitude error was employed as a performance metric for the DNN.
Details of the calculation are described in the \textbf{Methods} section.

Regarding the generation of the training dataset, it is worth noting that the DNN inherently possesses the ability to extract the amplitude weights of individual eigenmodes with the training of pure eigenmode data.
Unlike traditional deep learning-based methods on extracting amplitude weights~\cite{an2020deep,zhang2022learning}, a large dataset is not required, since the DNN is a physics-embedded network architecture. The hypothesis is that if the DNN can demultiplex individual spatial eigenmodes, it can also demultiplex their superposed fields. In this context, a dataset containing \textit{N} eigenmodes and their corresponding target field distribution is sufficient.
%The target field consists of \textit{N} regions, each corresponding to a specific mode, where the intensity (i.e., the pixel value) within each region directly represents the associated amplitude weight.
Therefore, for a 6-mode MMF, only 6 pairs of data are generated, enabling a fast training process. In this work, a three-layer DNN is constructed.
The loss curve during the training process is shown in Figure~\ref{fig:results_simulation_eigenmodes_6modes}(a). The network is trained for 1000 epochs in total, which takes approximately 10 seconds. A closer inspection of the curve reveals that the network almost converges within the first 100 epochs. After the training, the phase masks of the three-layer DNN are optimized for demultiplexing the intended MMF.  Figure~\ref{fig:results_simulation_eigenmodes_6modes}(b) shows the phase profile of each layer. Each pixel can modulate the phase of the propagating light from 0 to $2\pi$. In the experiment, each pixel is fabricated as a refractive cuboid with a depth tailored to the designed operating wavelength.

We first evaluated the DNN using pure eigenmode data. Taking the $LP_{11a}$ mode as an example, the simulated optical light propagation process, see Figure~\ref{fig:results_simulation_eigenmodes_6modes}(c), shows that the field is gradually modulated through the three diffractive layers and ultimately concentrated in the designated detection region. Cross-sectional field profiles confirm that the optical energy remains well confined along the propagation path, with negligible leakage.
All six eigenmodes were subsequently tested, and the corresponding output distributions are shown in Figure~\ref{fig:results_simulation_eigenmodes_6modes}(d). 
In each case, the DNN achieves clear spatial separation of the modes, with more than $95\%$ of the optical energy directed to the correct output region as given in Figure~\ref{fig:results_simulation_eigenmodes_6modes}(e). 
We further assessed the impact of network depth on demultiplexing performance.
Increasing the number of diffractive layers yields improved mode isolation.
The mean mode isolation increases from \SI{6.09}{\dB} to \SI{15.71}{\dB} as the number of layers increases from two to three, with the largest improvement observed between two and three layers. Increasing the depth to 5 yields a mode isolation of \SI{22.71}{\dB}.
To further investigate the DNN's feasibility, we also investigated the influence of the operation wavelength. Although the DNN is optimized for operation at 1568 nm, it demonstrates strong spectral robustness, maintaining mode isolation above \SI{13}{\dB} across the 1540–1595 nm wavelength range, as shown in Figure~\ref{fig:results_simulation_eigenmodes_6modes}(f). This result indicates that the bandwidth is large enough to cover multiple communication bands. Although slight differences exist among individual modes, their overall trends remain consistent.
Collectively, these results confirm that the DNN can reliably demultiplex spatial eigenmodes.

\begin{figure}[tbp]
    \centering
    \includegraphics[width=0.86\textwidth]{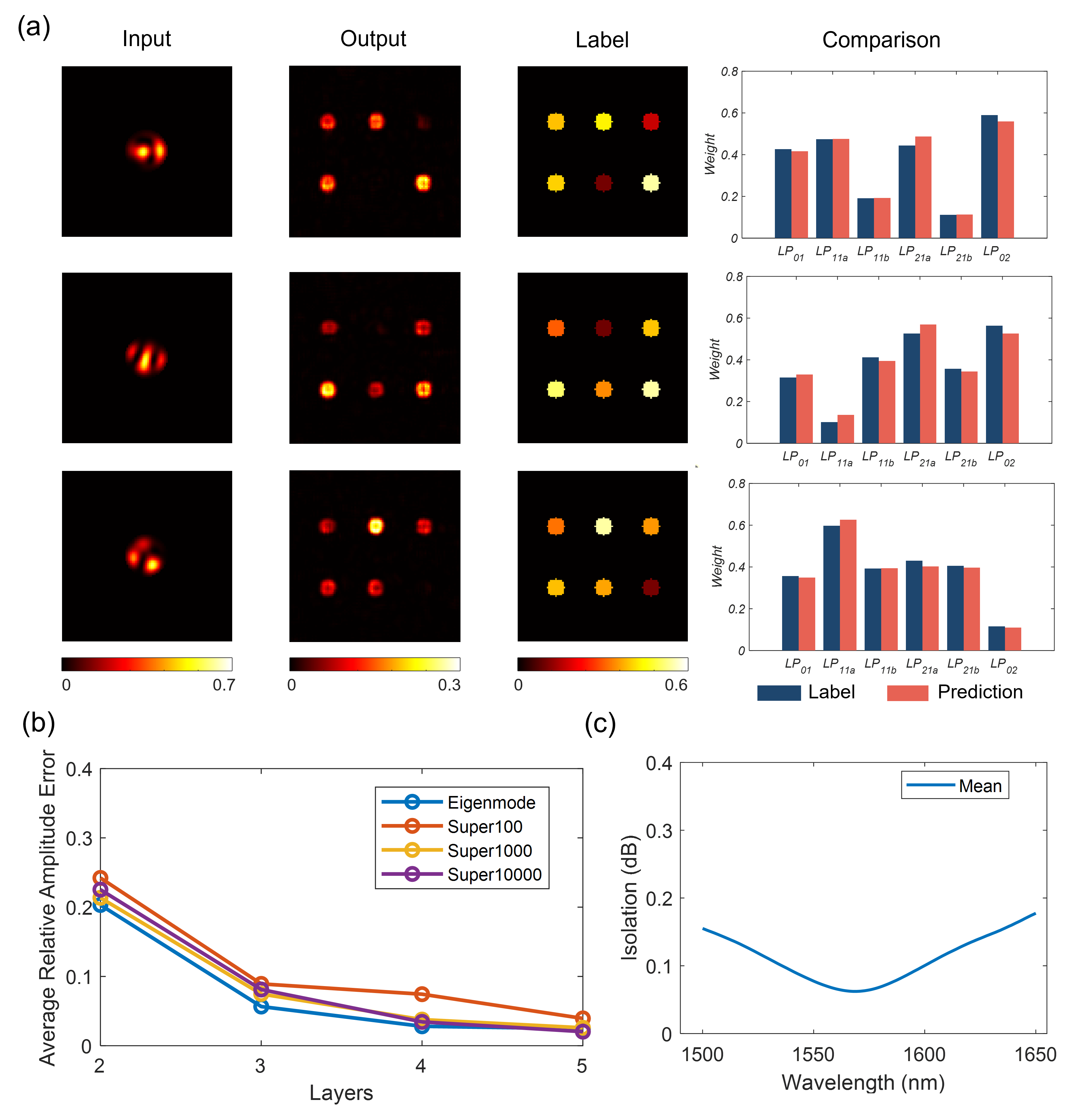}
  \caption[Visualization of the light propagation inside DNN for mode classification.]{
   Performance evaluation of DNN on synthetic superposed data. (a) 3 different superposed light fields are used as inputs to the DNN. The output of the DNN and the corresponding label are also given. The bar chart depicts the amplitude difference across different modes.
   (b) Relative amplitude weights error of the DNN trained with different datasets under different DNN depths. (c) Relative amplitude weight error across different optical wavelengths.
  }
  \label{fig:results_simulation_eigenmodes_6modes_super}
\end{figure}

Afterwards, we evaluated the DNN using randomly superposed modal fields.
To emulate arbitrary fiber speckle patterns, the relative modal phases were uniformly sampled in $(0, 2\pi)$, producing complex field distributions representative of practical propagation conditions.
Several test examples and their corresponding DNN outputs are shown in Figure~\ref{fig:results_simulation_eigenmodes_6modes_super}(a).
The diffractive layers reshape the input speckle field into spatially separated  spots, with each mode focused within its designated detection region.
The modal amplitude weights are then extracted by taking the square root of the detected power in each region, followed by L2-normalization.
The comparison between the predicted weights and the ground truth confirms that DNN separates the contribution of each mode well. 
A total of 1000 superposed modal fields were tested. The DNN achieves an average amplitude error of $0.0283$ and a relative amplitude weights error of $6.93 \%$. These results confirm high accuracy in the measurement of amplitude weights.
We further tested the DNN with identical amplitude weights but varying relative phase weights among modes. As illustrated in \textbf{Figure S3}, the DNN remains capable of accurately recovering the amplitude weights, demonstrating robustness against phase variations. 
However, it should be noted that the DNN cannot operate without phase information. If the input to the DNN contains only the amplitude distribution and the phase is set to zero, the network produces incorrect modal weights (see \textbf{Figure S4}). In this case, the energy is primarily concentrated in the mode $LP_{01}$, since it possesses a uniform (plane) phase distribution.
Collectively, these results demonstrate that, although trained only on individual eigenmodes, the DNN generalizes effectively to arbitrary superposed MMF fields.

% The amplitude error and the relative amplitude weights error are calculated to evaluate the performance of the DNN. 

To compare the two training strategies, we additionally generated pseudo-random mode-combination (PRMC) datasets with varying sizes. For both approaches, training with pure eigenmodes or with PRMC samples, the number of training epochs was fixed at 1000, and 1000 test samples were used for evaluation.
As summarized in Figure~\ref{fig:results_simulation_eigenmodes_6modes_super}(b), training exclusively with pure eigenmodes yields superior or comparable performance to that obtained using PRMC data. The small observed deviations are attributed to stochastic variations during optimization.
Notably, when only a small set of PRMC samples is used, the DNN performs worse than the model trained solely on eigenmodes. This degradation stems from the limited representativeness of small PRMC datasets, where certain modes may be insufficiently sampled. 
Although performance improves as more PRMC data are introduced, the required training time increases accordingly.
As the number of supported modes grows, the advantage of training with pure eigenmodes becomes more pronounced.
Furthermore, We trained DNNs with varying depths. The relative amplitude error with various layers is shown in Figure~\ref{fig:results_simulation_eigenmodes_6modes_super}(b). Adding diffractive layers benefits both training with PRMC data and eigenmodes data. The result indicates that more accurate demultiplexing can be achieved with more layers. 
Nevertheless, this does not mean that increasing the number of layers continuously improves the results. For six modes, the performance improvement gained from deeper DNN progressively decreases. Furthermore, adding more layers results in further loss of optical energy. 
% underscoring its efficiency and scalability for high-dimensional mode-decomposition systems.
Additionally, the performance of DNN over broadband wavelengths has been tested. Although the DNN is optimized for operation at 1568 nm, it demonstrates strong spectral robustness, maintaining relative error below $10\%$ across the 1527–1605 nm wavelength range, as shown in Figure~\ref{fig:results_simulation_eigenmodes_6modes_super}(c).

% Although slight differences exist, they are primarily attributed to the stochastic variations in the training process.
% It is worth noting that when only few PRMC samples are used for training, the DNN performance is actually inferior to that achieved using only eigenmodes. This is primarily attributed to the limited representativeness of such a small dataset, in which certain modes may be underrepresented. While increasing the amount of training data generally improves performance, whereas the training time increases as well.
% As the number of modes increases, the advantages of training with pure eigenmodes become increasingly evident. 

% \begin{table}[htbp]
% \centering
% \caption{Performance comparison for different types of training data (6 modes\textcolor{red}{to be changed}).}
% \begin{tabular}{|l|c|c|c|c|}
% \hline
% Training data type &$\Delta \rho$ & $\Delta \rho_r$ &$\bar{\Gamma}_{amp}$ & $\sigma_{\Gamma}$ \\ \hline

% Eigenmodes & 0.0825 &  0.1429      & 0.9x04 & 0.0085  \\ \hline
% *$PRMC_{10}$  & 0.0876 & 0.1517    & 0.9x19 & 0.0067 \\ \hline
% $PRMC_{100}$       & 0.0788 & 0.1365  &0.9x13 & 0.0070 \\ \hline
% $PRMC_{1000}$       & 0.0674 & 0.1167   & 0.9x29 & 0.0066 \\ 
% \hline
% \end{tabular}
% \footnotesize
% \label{tab:DNN_MD_amplitude_training_data_comparison}
% \begin{tablenotes}
% \small
% \item *$PRMC_n$ denotes PRMC training data with $n$ samples.
% \end{tablenotes}
% \end{table}

% we did the same test for more modes in the simulation....

\subsection*{Fabrication of DNN using 3D-printed technique}

\begin{figure}[t!]
    \centering
    \includegraphics[width=0.86\textwidth]{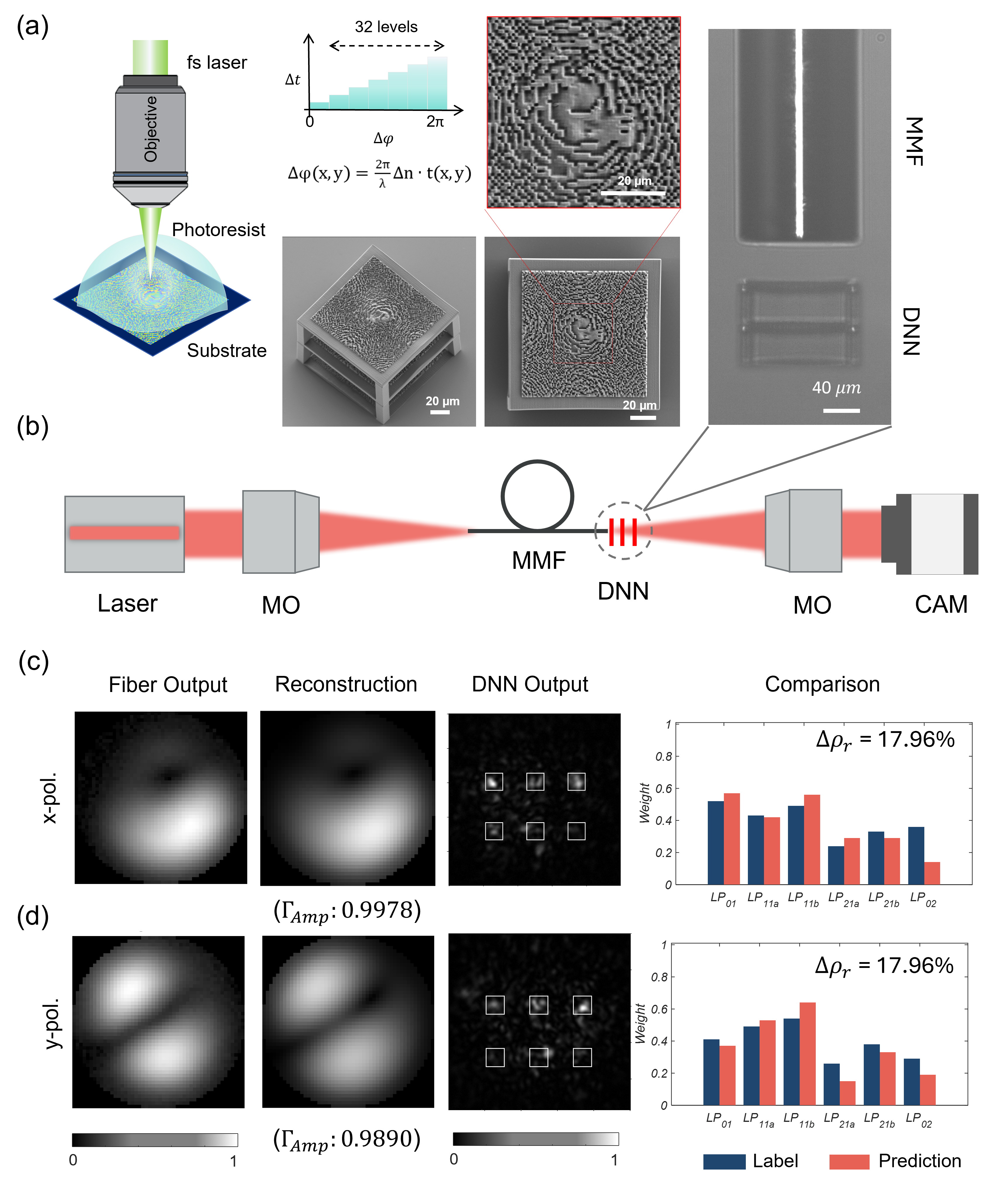}
  \caption[Example of using 3D printed DNN for performing amplitude decomposition in the experiment.]{ Experimental results of using 3D printed DNN.
  (a) Experimental fabrication of the DNN chip for MMF modes demultiplexing and detection. SEM images of the three-layer DNN and the microscopy image of MMF and DNN are shown.  
  (b) A schematic diagram of the experimental setup. Abberations: MO, microscope objective, CAM. camera. 
  (c),(d) Experimental results of two mutually perpendicular polarizations. The intensity value within each detection area represents the power ratio among different modes. The amplitude weights are then obtained by taking the square root of the intensity value. Each sample is independently normalized. }
  \label{fig:DNN_md_exp_result}
\end{figure}

The fabrication process of the DNN chip is achieved using a home-built two-photon nanolithographic (TPN) system. After the phase information of the DNN are calculated, these phase plates carrying the correct phase information are designed to impart a spatially varying phase shift on an incident optical wavefront. In our TPN fabrication system, this is achieved by locally modulating the thickness or refractive index of a polymerized 3D microstructure. Because TPP is a voxel-by-voxel fabrication process based on nonlinear absorption, arbitrary phase patterns with sub-micron feature sizes can be realized. The induced phase modulation at a given pixel location (x,y) is determined by the local optical path length (OPL): $\Delta\phi(x,y) = \frac{2\pi}{\lambda}\,\Delta n \cdot t(x,y)$, where $\Delta n$ is the refractive index contrast between polymerized and unpolymerized regions, and t(x,y) is the local structure thickness, and the calculated phase (thickness) are divided into 32 phase levels for the ease of the fabrication. 
The fabrication process and the fabrication result is schematically shown in Figure~\ref{fig:DNN_md_exp_result}(a). The scanning electron microscopic (SEM) images of the three-layer DNN chip for MMF modes detection is shown on the right of Figure~\ref{fig:DNN_md_exp_result}(a). The phase modulation layers are fabricated separately on a scaffold with the size of \SI{120}{\micro\metre} by \SI{120}{\micro\metre} by \SI{80}{\micro\metre}, and the layers are accurately aligned with a layer distance of \SI{40}{\micro\metre}.

% \begin{figure}[ht]
%     \centering
%     \includegraphics[width=0.86\textwidth]{figures/figure_fabrication.png}
%   \caption[Example of using 3D printed DNN for performing amplitude decomposition in the experiment.]{Experimental fabrication of the DNN chip for MMF modes demultiplexing and detection. }
%   \label{fig:fabrication}
% \end{figure}

\subsection*{Experimental validation using 3D-printed technique}

To experimentally verify the performance of the proposed DNN, we constructed an optical setup to test the demultiplexing capability under practical multimode excitation conditions. The experimental system is illustrated in Figure~\ref{fig:DNN_md_exp_result}(b).
A tunable continuous-wave laser at a wavelength of \SI{1568}{\nano\metre} is coupled into a MMF with a core radius of \SI{25}{\micro\metre} and NA with a value of 0.1. The fiber supports six modes at the operation wavelength. The fiber output is first imaged onto an infrared camera. Since the MMF output contains mixed polarization states, a polarizer is placed in front of the camera to select a single polarization state. Before training and fabricating the DNN, the first step is to characterize the profile of the MMF.
Therefore, we collected 100 output samples from the fiber and employed a well-trained deep neural network (MTNet)~\cite{zhang2022learning} to perform MD to characterize the profiles of the eigenmodes, which are the basis for training the DNN. Details of the characterization procedure can be found in the \textbf{Supplementary Materials}. 
After training and fabricating the DNN, it was placed directly after the MMF for experimental testing. The relative position between the fiber facet and the DNN was precisely adjusted using microstages. To ensure accurate alignment in the setup, an auxiliary 4-f imaging system was implemented to monitor and secure the relative distance between the fiber facet and the DNN. 

We first collected the output patterns from the MMF, see Figure~\ref{fig:DNN_md_exp_result}(c). The amplitude distribution of output is then obtained from the intensity image, which was cropped and resized to meet the requirements of MTNet~\cite{zhang2022learning}. After pre-processing, the MTNet yielded the modal amplitude weights and relative phase weights, which were used to reconstruct the light field. 
Different from the typical phase-type modulator-based implementation~\cite{fontaine2019laguerre}, which can only manipulate one polarization, the 3D-printed DNN can simultaneously demultiplex optical fields from both polarization states, since the diffractive structure modulates light independent of polarization.
We repeated the measurement for another orthogonal polarization by rotating the polarizer $90^{\circ}$.
As shown in Figure~\ref{fig:DNN_md_exp_result}(c) and (d), the reconstructed fields achieved high correlation coefficients of about $99\%$ with the original fields, confirming the accuracy of the extracted modal weights. These weights serve as a reference (ground truth) for evaluating the demultiplexed results by DNN.
Six detection areas were localized as shown in Figure~\ref{fig:DNN_md_exp_result}(c). The optical energy within each region was used to measure the modal power ratio. The amplitude weights of each mode are then obtained by calculating the square root of the power. The relative weight errors of the presented examples are both about $18\%$. 
%Across \textbf{X} experimental test samples, the DNN achieved an average relative error of approximately \textbf{$XX\%$}, 
Experimental results were slightly worse than the simulation results. This discrepancy mainly arises from two factors. First, imperfections in the experimental setup lead to alignment errors between the DNN and the fiber facet. Second, fabrication imperfections and the discretized phase levels of the printed diffractive layers introduce additional deviations from the ideal model.
We also evaluated the insertion loss under the single-polarization condition. The insertion loss shows slight variations with different inputs, with a minimum value of -4.7 dB and an average of -5.5 dB.
% This demonstrates that the fabricated DNN possesses inherent polarization tolerance, enabling direct processing of realistic multimode-fiber outputs without polarization pre-selection. 

% Although our device was optimized at 1550 nm, its performance within the bandwidth of 1540–1560 nm is relatively stable. For all modes, the mode fidelity varies less than 1 dB, and the crosstalk stays below 15 dB for most of the bandwidth. This range falls within dense WDM systems, which indicates that our device does have WDM compatibility.
%https://pubs.acs.org/doi/10.1021/acsphotonics.1c01744

% Topical subheadings are allowed.

% \section*{Discussion}

% The Discussion should be succinct and must not contain subheadings.
\section*{Discussion}

% \textbf{The same tests have also been conducted on more modes, see Supplementary Materials.} 

% ...definition of the error should be presented ...is it fluctuation or systematic error... how it is depending on SNR....please cite this paper. the error/uncertainty is presented in depencency of SNR....

In this work, we proposed and experimentally demonstrated a compact, high-speed mode demultiplexer based on a DNN.
The device enables all-optical demultiplexing in a six-mode MMF, demonstrating its potential in next-generation optical communication systems exploiting space division multiplexing.
Unlike SLM- or metasurface-based demultiplexers, 
%which require bulky optics and nanometer-scale alignment, 
the presented approach realizes the first chip-scale mode demultiplexer fabricated via two-photon 3D nanolithography. 
The DNN was designed and optimized in simulation using a physics-informed deep-learning framework. Although the training uses only a small set of eigenmode data, the DNN exhibits strong generalization capability and is able to demultiplex arbitrary optical fields within the modal domain. 
A mode isolation of 15.71 dB was achieved for pure eigenmodes, confirming the network’s capability for effective mode separation.
For randomly superposed fields, the relative amplitude-weight error was approximately $6.93\%$. A smaller error indicates a higher precision in resolving the modal components, thereby offering an accurate reference for closed-loop control schemes or DSP-assisted signal correction~\cite{weng2019recent}, which is crucial for commercial MMF-based networks as the mode conversion is inevitable. It should be noted that the DNN achieves slightly higher accuracy on pure eigenmodes compared with arbitrarily superposed optical fields.
This degradation primarily arises from two factors: the training data included only eigenmode samples, leading to slight overfitting. The second is the limited spatial resolution of the printed DNN, which restricts its ability to distinguish subtle field variations in complex superpositions.
Investigation using synthetic data shows that increasing the number of layers can further reduce this modal weights error (see Figure~\ref{fig:results_simulation_eigenmodes_6modes_super}(b)). However, excessive depth introduces optical losses and fabrication accumulation errors.
A balance between achievable performance and layer number is therefore essential. 

The experimental implementation confirmed the feasibility of the proposed fully-optical demultiplexer, with only a slight performance drop attributed to alignment tolerances and printing imperfections.
Importantly, the DNN was trained solely with eigenmodes yet effectively demultiplexed arbitrary superposed fields, significantly simplifying the training process and reducing computational requirements.
Compared with digital neural network-based methods such as MTNet~\cite{zhang2022learning}, which require tens of thousands of samples and hours of training, the DNN converges within seconds on a standard GPU, enabling rapid adaptation to dynamic optical environments.
The fabricated DNN chip demonstrates broadband operation and can, in principle, be tailored for narrowband conditions to integrate wavelength- and space-division multiplexing within a single device~\cite{li2023massively}.
Furthermore, the device operates with dual-polarization input using only a single PBS, showing the feasibility for polarization-division multiplexing systems. By including digital optical phase conjugation~\cite{buttner2020velocity, czarske2016transmission} or transmission matrix such method~\cite{gomes2022near} into the calibration of the system, high fidelity mode transmission can be achieved as the input of DNN. 
Thanks to its compact footprint and passive nature, the 3D-printed DNN can be directly integrated onto the fiber facet, eliminating alignment constraints and paving the way toward scalable on-fiber or on-chip photonic neural processors. When integrated with dynamic optical elements serving as a mode multiplexer, the proposed DNN enables a fully closed-loop and distortion-free multimode communication system~\cite{pohle2023intelligent,rothe2024unlocking}.
Despite its preliminary nature, this study highlights the strong potential of 3D-printed diffractive neural networks for real-time, low-power, and highly parallel optical signal processing in modern multimode communication systems.

%cite~\cite{zhao20243d}

% Furthermore, maybe due to AI more flexible for different fibers
% +
% scaling to higher mode numbers, correct...what is your opinion?

% In this paper, the DNN is used for the first time worldwide for the MD of an MMF for distortion-free information transmission,
% where mode conversions can be measured and corrected or exploited
% ...explain how distortion-free transmission is achieved.........

% \textbf{Similar evaluations were performed for scenarios involving 3 and 10 modes, see Supplementary}. 

% The scalability.....  for 10 modes, 55 modes.... 

% XXX

% \textbf{demultiplexing also related to phase distribution.....}

% Adaptive optical components and wavefront shaping methods promise to control such scattering in complex media~\cite{vellekoop2007focusing,popoff2010measuring,papadopoulos2012focusing,ploschner2015seeing,    carpenter2016complete,turtaev2018high,valencia2020unscrambling }.

\section*{Methods}

\subsection*{3D GS-TPN}

The printing of DNN was achieved using a 3D GS-TPN fabrication system built in-house that incorporates a femtosecond laser source (Coherent Axon 1064), second-harmonic generation module (Coherent SGH), piezo nanotranslation stage (PI-545), one pair of two-dimensional galvo mirrors (Thorlabs GVS002), scan lens (Thorlabs SL50-CLS2) and tube lens (Thorlabs TTL200MP) utilized to reduce the focus aberration during the galvo-scanning fabrication process. The mechanical and dynamic control of the system is controlled by a home-developed program, by which the 3D CAD models can be translated into the fabrication documents required by the program. The galvo mirrors rotate at a fast speed without the movement of the nanopiezo translation stage, thus enabling ultrafast movement of the fabrication voxel.

% \subsection*{ Experimental setup}

% A coherent, broadband continuous wave laser was used 
% (ID Photonics CoBrite DX1)

% fiber 
% Thorlabs, M68L02

% polarizer 
% XXX 

% two objectives, 

\subsection*{Hyperparameters of training}

The DNN was trained using the PyTorch deep-learning framework (Python 3.9.24, PyTorch 2.8).
All simulations and training procedures were executed on a computer equipped with an AMD EPYC 7452 CPU and an NVIDIA RTX A6000 GPU.
The network was optimized using the Adam optimizer with a mean-squared-error (MSE) loss function. The learning rate was set to $1e-3$ and decayed by 0.1 during training every 100 epochs.

\subsection*{Performance metrics for DNN}
To quantify the separation performance of the DNN, the mode isolation is defined as the power ratio between the desired mode channel and the sum of all undesired mode channels when a single input mode is launched.
The average mode isolation over all modes is given by
\[
\mathrm{MI}
= \frac{1}{N} \sum_{n=1}^{N} 10\log_{10}\!\left(
\frac{P_{n}}{P_{all}}
\right),
\]
where $P_n$ denotes the normalized optical power detected in the target region for the \(n\)-th mode, and $P_{all}$ is the total power detected in all other regions.

The insertion loss is as
 \[
 \mathrm{IL}
 = -10\log_{10}(\frac{P_{\mathrm{out}}}{P_{\mathrm{in}}}),
 \]
where $P_{in}$ and $P_{out}$ represent the optical power before and after the DNN, respectively.

The weight error between the predicted mode weights and the actual mode weights is calculated as $\Delta\rho=\left|\rho_p-\rho_t\right|$.
Here, the subscript \textit{p} indicates a predicted value, and the subscript \textit{t} denotes a target value. $\Delta\rho$ refers to the amplitude error. In order to provide a more comprehensive response to the accuracy of the predicted weights for different modes, relative amplitude error was calculated as $\Delta\rho_r=\langle\left|\rho_p-\rho_t\right|\rangle / \langle ave(\rho_t)\rangle$, where the $ave(\rho_t)$ is the average value of the mode weights. As the number of modes increases, the relative amplitude error becomes a more reliable indicator of prediction accuracy.

\bibliography{sample}

\section*{Acknowledgements}

%Acknowledgements should be brief, and should not include thanks to anonymous referees and editors, or effusive comments. Grant or contribution numbers may be acknowledged.

We would like to express our sincere gratitude to the German Research Foundation for funding the Reinhart Koselleck project for highly innovative research (CZ 55/61-1, project number: 560574412). This research was also funded by the German Federal Ministry of Education and Research within the framework of the 6G-life (funding code: 16KISK001K) and QUIET (project code: 16KISQ092) projects.

\section*{Author contributions statement}

% Must include all authors, identified by initials, for example:
% A.A. conceived the experiment(s),  A.A. and B.A. conducted the experiment(s), C.A. and D.A. analysed the results.  
Qian Zhang and Juergen Czarske conceived the concept. 
Haoyi Yu and Meng Chao conducted the experimental fabrication of the DNNs chip.
Jie Zhang and Haoyi Yu conducted the experimental characterization of the DNNs chip.
Qian Zhang, Yuedi Zhang, and Yu Miao conducted the programming and the training of the DNNs.
Jiali Sun trained the MTNet for decomposing fiber output. 
Qian Zhang, Jie Zhang, Haoyi Yu, and Juergen Czarske analysed the results.
Juergen Czarske, Min Gu, and Qiming Zhang supervised the project.
All authors participated in discussions and contributed to the writing of the paper. All authors reviewed the manuscript.

\end{document}